\documentclass[preprint,showpacs,preprintnumbers,amsmath,amssymb]{revtex4}

\usepackage{epsfig}
\usepackage{graphicx}
\usepackage{dcolumn}
\usepackage{bm}
\usepackage{url}
\usepackage[dvipsnames]{xcolor}
\usepackage[colorinlistoftodos,prependcaption,textsize=small]{todonotes}

\begin{document}



\title{Synergy as a warning sign of transitions: the case of the two-dimensional Ising model}

\author{
D. Marinazzo$^{1}$, L. Angelini$^{2,3}$, M. Pellicoro$^{2}$, S. Stramaglia$^{2,3}$}

\affiliation{$^1$ Department of Data Analysis, Ghent University, 2 Henri Dunantlaan, 9000 Ghent, Belgium\\}\affiliation{$^2$ Dipartimento Interateneo di Fisica,
Universit\'a degli Studi Aldo Moro, Bari and INFN, Sezione di Bari, via Orabona 4, 70126
Bari, Italy\\} \affiliation{$^3$ Center of Innovative Technologies for Signal Detection and Processing (TIRES),Universit\'a degli Studi Aldo Moro, Bari, via Orabona 4, 70126
Bari, Italy\\}

\date{\today}

\begin{abstract}
We consider the formalism of information decomposition of target effects from multi-source interactions, i.e. the problem of defining redundant and synergistic components of the information that a set of source variables provides about a target, and apply it to the two-dimensional Ising model as a paradigm of a critically transitioning system. 
Intuitively, synergy is the information about the target variable that is uniquely obtained taking the sources together, but not considering them alone; redundancy is the information which is shared by the sources. To disentangle the components of the information both at the static level and at the dynamical one, the decomposition is applied respectively to the mutual information and to the transfer entropy between a given spin, the target, and a pair of neighbouring spins (taken as the drivers).
We show that a key signature of an impending phase transition (approached from the disordered size) is the fact that the synergy peaks in the disordered phase, both in the static and in the dynamic case: the synergy can thus be considered a precursor of the transition. The redundancy, instead, reaches its maximum at the critical temperature. The peak of the synergy of the transfer entropy is far more pronounced than those of the static mutual information. We show that these results are robust w.r.t. the details of the information decomposition approach, as we find the same results using two different methods; moreover, w.r.t.  previous literature  rooted on the notion of Global Transfer Entropy,  our results demonstrate that considering as few as three variables is sufficient to construct a precursor of the transition, and provide a paradigm  for the investigation of a variety of systems prone to crisis, like financial markets, social media, or epileptic seizures.

\end{abstract}\maketitle

\maketitle
High dimensional physical, biological or social systems are formed by a large number of interacting components, and their macroscopic evolution is the product of intertwined mechanisms at the lower scale. These systems are typically characterized by stochastic disordered dynamics, and sudden transitions between different regimes. In this framework it is of crucial importance to develop methods that  are able to identify precursors, i.e. warning signals, to forecast the transition before it takes place, notably in processes such as financial market crisis, polarization of news, or epileptic seizures.  Since the mechanisms underlying the rapid rearrangement of the dynamics, connected with the transition, may be different in different systems in the microscopic details but similar at the systemic level, it is important to find and assess new paradigms which can act as a conceptual background for a variety of systems prone to crisis. 
 Using the temperature decrease of a spin system as a model, a precursor marker of a dynamical transitions from disorder to order has been proposed \cite{barnettprl2013}: the global transfer entropy (GTE), rooted in the formalism of  information dynamics.
The transfer entropy (TE) is a quantity introduced in \cite{schreiber} and based on appropriate conditioning of transition probabilities, see \cite{book_te} and references therein; it is able to effectively distinguish driving and responding elements, and to detect asymmetry. Let $s_i$ be a given spin, taken as the target variable, and let us denote $\tilde{s}_i$ its value at the next time step. Let $\{s_j s_k \cdots s_r\}$ be a group of spins which is assumed to be candidate driver for the spin $s_i$. The transfer entropy from this group of drivers to the target is defined in terms of the following conditional mutual information:
\begin{equation}
    T_{j k\cdots r \to i}= I\left(\tilde{s}_i ;\{s_j s_k \cdots s_r\}|s_i\right).
    \label{tegeneral}
\end{equation}
For each target variable, GTE is computed by measuring the information provided by all the other variables about the future of each target, i.e. taking as driving group all the variables at hand but $s_i$, eventually averaging over all the possible targets. The pairwise TE, at fixed target, corresponds instead to the amount of information provided by each other variable, $I\left(\tilde{s}_i ;s_j |s_i\right)$, averaged over the driving variable $s_j$. 

For a 2D lattice Ising model with Glauber dynamics, information flow, as quantified by GTE, attains a maximum strictly in the disordered (paramagnetic) phase and is thus able to predict an imminent transition \cite{barnettprl2013}. On the other hand, the pairwise TE and the mutual information peak at criticality; the gradients of all measures appear to diverge as the temperature  tends to the critical value.  
As discussed in \cite{prokopenko} GTE is a measure of collective information transfer, capturing both pairwise and higher-order (multivariate) correlations to a variable. As remarked in \cite{barnettprl2013}, its peak can be interpreted in terms of conflicting tendencies among these components as the  disorder decreases, while tending to the phase transition point: pairwise correlations can be expected to become more prevalent than higher-order multivariate effects as the critical temperature is approached. It follows that disentangling explicitly the components of the collective information flow is needed to get a better description of the system in the proximity of the transition. GTE has  also been shown to be a predictor of finite first order transitions, like in the canonical Potts spin model \cite{potts}. These findings have supported the conjecture that generically for systems featuring order-disorder phase transitions, a key signature of an impending phase transition (approached from the disordered size), is a peak in GTE. Recently GTE has been used to predict crash events of stock markets in \cite{stock}.

In order to calculate GTE from data (1) all the relevant variables of the system must be measured, and (2) one should have access to dynamical data (GTE evaluates dynamic dependencies based on lagged conditioned correlations). 
In a recent paper \cite{sootla} it has been shown that  conditions (1) and (2) are not necessary and that a precursor of crisis can be built even measuring a limited number of variables and using static (mutual information-based) quantities.
The key to answer this question resides indeed in the effort spent in recent years by many researchers to achieve a satisfactory formulation of the information decomposition of target effects from multi-source interactions, i.e. the problem of defining redundant (or shared), unique and synergistic (or complementary) components of the information that a set of source variables provides about a target, see \cite{ID} and references therein. Applying this formalism to the mutual information between a target spin and two of its neighbors in the two dimensional Ising model of size $128\times 128$, in \cite{sootla} it has been shown that the synergy peaks in the disordered phase considering as few as three variables; lagged correlations are not necessary to this scope. However, in \cite{sootla} a particular prescription introduced in \cite{MMI} was used to obtain the decomposition of the mutual information: what remained to be clarified is whether such phenomenon depends on the details of the information decomposition methodology applied. 
In this work on one side we show, simulating the 2d Ising model on a $512\times 512$ lattice, that the findings of \cite{sootla} do not depend on the way the information decomposition of the mutual information is done; indeed using a different method we find the same results as those from \cite{MMI}.  Moreover we also show that performing the decomposition of the transfer entropy from the two neighboring spins, to the target spin, leads to a far more pronounced peak of the synergy w.r.t. those of the mutual information in the paramagnetic phase, thus providing a clear connection between the information decomposition frame and the results about GTE described in \cite{barnettprl2013}.

Let us consider the two dimensional Ising model, where spins on a regular lattice are characterized by the Hamiltonian
\begin{equation}
    H=-\beta \sum_{\langle i j\rangle} s_i s_j,
    \label{hamiltonian}
\end{equation}
$\beta$ being the coupling and the sum being performed over nearest neighbor pairs of spins. This model shows a second order phase transition at $\beta_c \approx 0.4407$, in correspondence with long range correlations in the system \cite{baxter}. The mutual information of a pair of nearest neighbor spins $I\left(s_i;s_j\right)$ has been calculated in \cite{matsudo} and peaks at criticality; it represents the static information about $s_i$ contained in the stochastic variable $s_j$. Analogously we can consider the spin $s_i$ as a target, and spins $s_j$ and $s_k$ as two drivers.  The information on $s_i$ contained in this pair of drivers is the mutual information $I\left(s_i ;\{s_j s_k\}\right)$.
The desirable information decomposition is:
\begin{eqnarray*}
        I\left(s_i ;\{s_j s_k\}\right)=U^I_{j\to i}+U^I_{k\to i}+R^I_{j k\to i}+S^I_{j k\to i},\\
    I\left(s_i ;s_j\right)=U^I_{j\to i}+R^I_{j k\to i},\\
        I\left(s_i ;s_k\right)=U^I_{k\to i}+R^I_{j k\to i}.\\
        \label{exp:mi}
\end{eqnarray*}
In the expansion above, the terms $U^I_{j\to i}$ and $U^I_{k\to i}$ quantify the components of the information about the target $s_i$ which are unique to the sources $s_j$ and $s_k$, respectively, thus reflecting contributions to the predictability of the target that can be obtained from one of the sources when it's treated as the only driver, and not from the other source.  Each of these unique contributions sums up with the redundant information $R^I_{j k\to i}$ to yield the mutual information between one source and the target according to the classic Shannon information theory. Then, the term $S^I_{j k\to i}$ refers to the synergy between the two sources while they provide information about the target, intended as the information that is uniquely obtained taking the two sources $s_j$ and $s_k$ together, but not considering them alone. Since in the expansion four quantities are unknown and just three equations are at hand, the information decomposition in unique, redundant and synergistic parts is a missing piece in classical information theory. An additional ingredient to Shannon theory is needed to get a fourth defining equation  for providing an unambiguous definition of $U^I_{j\to i}$, $U^I_{k\to i}$,   $R^I_{j k\to i}$ and  $S^I_{j k\to i}$. While several information decomposition definitions have been proposed arising from different conceptual definitions of redundancy and synergy \cite{harder,griffith,quax}, in the case of three variables several of these definitions coincide. The so-called  minimum mutual information (MMI) PID \cite{barrett} assumes that redundancy
is given by the minimum of the information provided by each individual source to the target, and hence is independent of the correlation between sources; moreover synergy is the extra information contributed by the weaker source when the stronger source is known, and can either increase or decrease with correlation between sources. Another proposal is the maximum entropy-based redundancy measure of \cite{MMI}, which is defined as follows. Let us call $p(s_i,s_j,s_k)$ the joint probability distribution of the three spins, and consider the set of all the probability distributions $q(s_i,s_j,s_k)$ that preserve the bivariate marginals involving $s_i$, i.e. $q(s_i,s_j)=p(s_i,s_j)$ and $q(s_i,s_k)=p(s_i,s_k)$. It is then assumed that unique information and redundancy are invariant within this set, while synergy depends on the specific form of the trivariate joint distributions. Synergy is thus determined as the difference between the mutual information for the original $p(s_i,s_j,s_k)$ and the minimum within the set of $q$ distributions:
\begin{equation}
    S^I_{j k\to i}=I\left(s_i ;\{s_j s_k\}\right)-\mbox{MIN}_q \;\; I\left(s_i ;\{s_j s_k\}\right).
    \label{synergy}
\end{equation}
In \cite{barrett} it has been shown that in the case of Gaussian stochastic variables the  MMI and the maximum entropy-based approach are equivalent and provide the same decomposition. Although these two methods in general give different results, for the problem at hand we have verified that they provide the same decomposition.

The transfer entropy from the pair of spins $s_j$ and $s_k$ to the target $s_i$ is defined as follows, in terms of conditional mutual information: $T_{j k\to i}=I\left(\tilde{s}_i ;\{s_j s_k\}|s_i\right)$, where $\tilde{s}_i$ is the future value of $s_i$ obtained when the system is updated with Glauber dynamics, like in \cite{barnettprl2013}. The pairwise transfer entropies are defined as $T_{j \to i}=I\left(\tilde{s}_i ;s_j|s_i\right)$ and $T_{k \to i}=I\left(\tilde{s}_i ;s_k|s_i\right)$, respectively. Note that conditioning on the past of the target $s_i$ (a step that in the information decomposition framework also includes synergies produced by the source with past of the target) removes shared information due to common history.
The information decomposition now reads:
\begin{eqnarray*}
        T_{j k\to i}=U^T_{j\to i}+U^T_{k\to i}+R^T_{j k\to i}+S^T_{j k\to i},\\
    T_{j \to i}=U^T_{j\to i}+R^T_{j k\to i},\\
       T_{k\to i}=U^T_{k\to i}+R^T_{j k\to i}.\\
        \label{exp:te}
\end{eqnarray*}
Also for the decomposition of the transfer entropy in this Ising system, we have verified that the adoption of the  MMI approach and the maximum entropy-based approach provide the same decomposition.  \cite{barrett}.
\begin{figure}[ht!]
\begin{center}
\includegraphics[width=7cm]{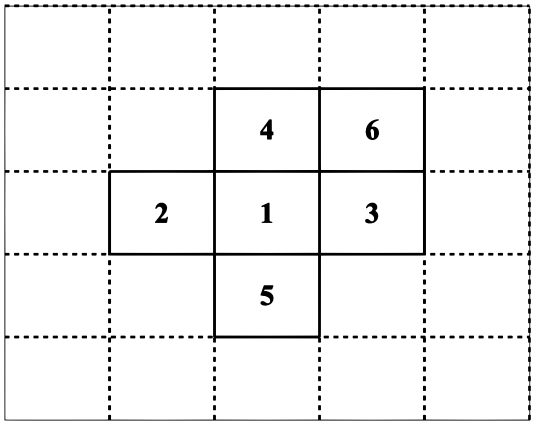}\caption{{\rm
The spatial arrangement of spins around spin $s_1$, to which the following figures refer.
 \label{fig1}}}
 \end{center}
 \end{figure}

We  simulated  a kinetic  Ising  model  of  size $N=L\times L$ with $L=512$ and periodic boundary conditions. Each update consisted of N (potential) spin-flips according  to  the  Glauber  transition  probabilities. Simulations  were  initially  run  for
a relaxation time of $10^4$ updates and statistics then collected over  a  further  $10^5$
updates. This  procedure  was  performed  for  1000  runs at  each  of  50  temperature  points  enclosing  the  phase transition. We will refer to the spin configuration depicted in figure (\ref{fig1}), where 1 is the target and we take two of the neighboring spins as the drivers.

\begin{figure}[ht!]
\begin{center}
\includegraphics[width=10cm]{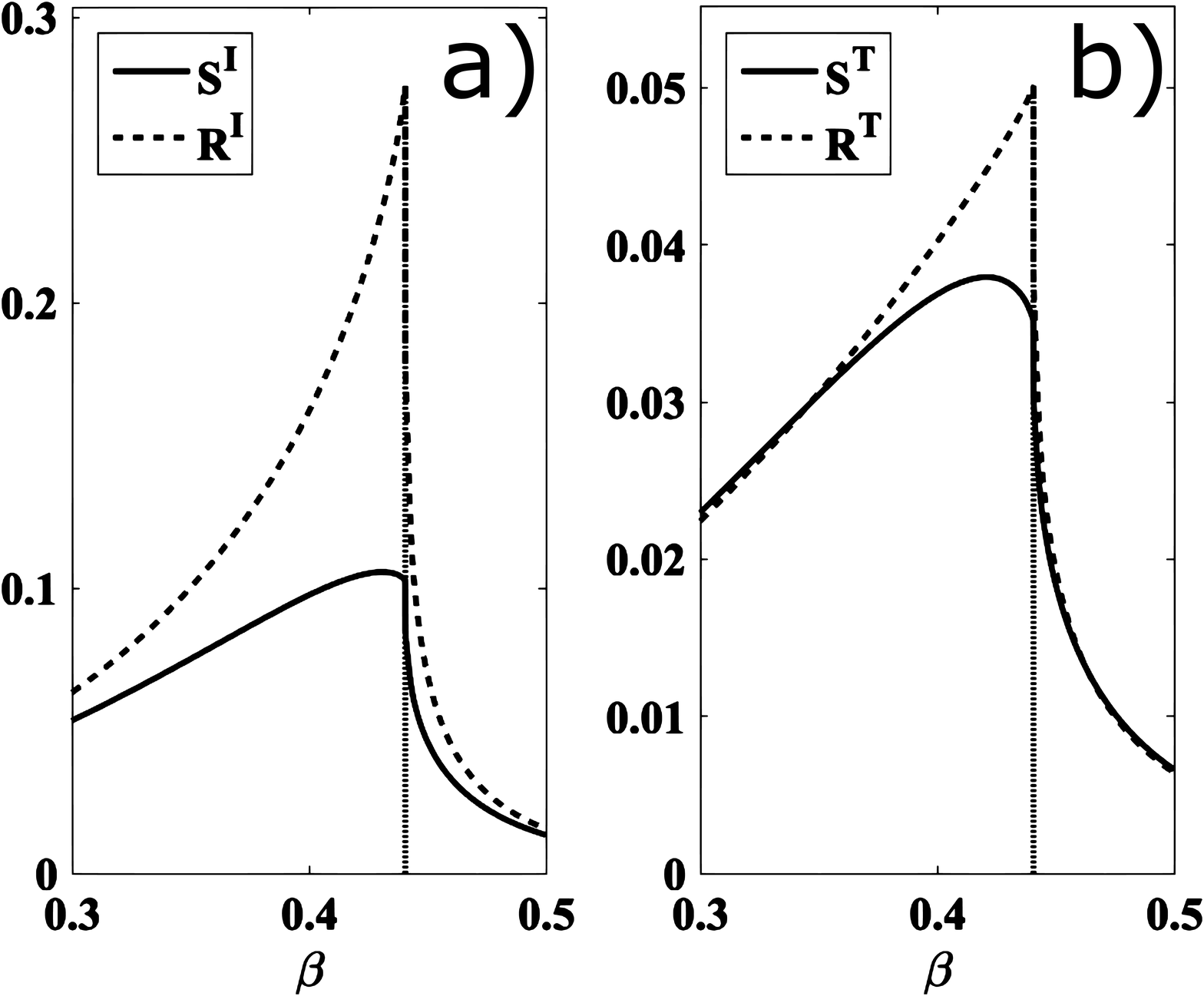}\caption{{\rm (a) The redundancy and synergy of spins $s_2$ and $s_3$ according to the decomposition of the mutual information, as a function of the coupling $\beta$. (b) The redundancy and synergy of spins $s_2$ and $s_3$ according to the decomposition of the transfer entropy, as a function of the coupling $\beta$.
\label{fig2}}}
\end{center}
\end{figure}

\begin{figure}[ht!]
 \begin{center}
\includegraphics[width=10cm]{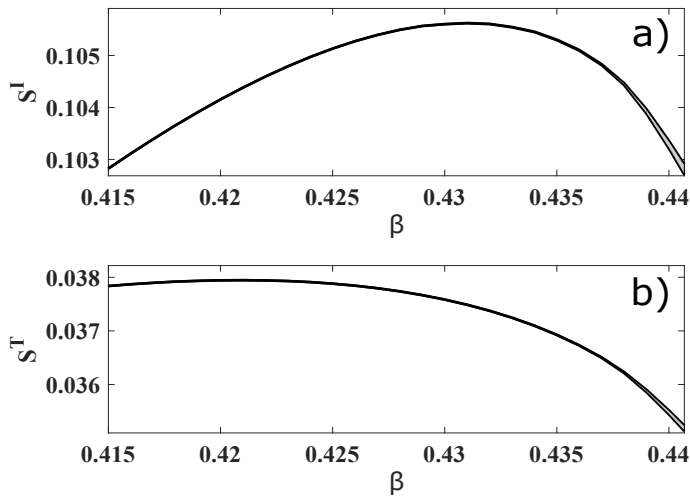}\caption{{\rm (a) The peak of the synergy in figure (\ref{fig2}a), related to the decomposition of the mutual information, is zoomed in.  (b) The peak of the synergy in figure (\ref{fig2}b), related to the decomposition of the transfer entropy, is zoomed in. The shaded areas, corresponding to the 95$\%$ confidence interval, is hardly visible. 
\label{fig5}}}
\end{center}
\end{figure}

Let us start considering the drivers $s_2$ and $s_3$, i.e. two opposite nearest neighbors to the target $s_i$. In this case the unique information terms are all vanishing, both for the static (mutual information) and the dynamic (transfer entropy) information decomposition: this property holds whenever the marginals corresponding to the two drivers are the same. In figure (\ref{fig2}) we depict the redundancy and the synergy as a function of the coupling $\beta$: while the redundancy peaks at criticality, the synergy of the two spins have a peak in the paramagnetic phase. In figure (\ref{fig5}) the peaks of the static and dynamic synergies have been zoomed in (the confidence interval is also shown but it is so narrow that it can be seen with much difficulty).  

\begin{figure}[ht!]
\includegraphics[width=10cm]{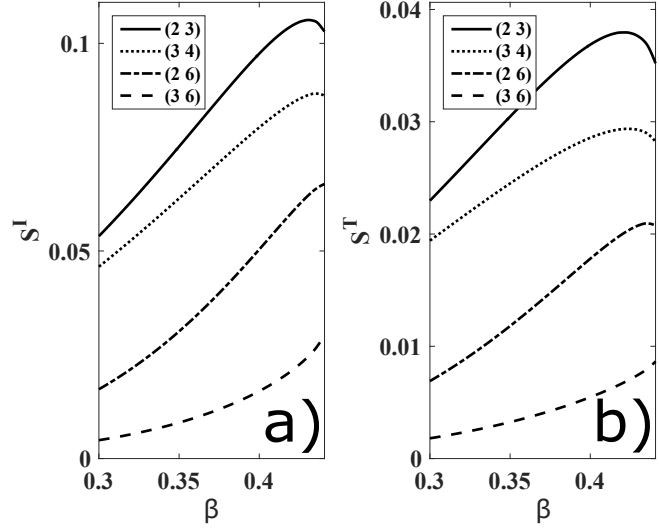}\caption{{\rm (a) The synergy of four pairs of driving spins (2-3, 3-4, 2-6, and 3-6) is depicted versus the coupling $\beta$, for the decomposition of the mutual information. (b) The synergy of four pairs of driving spins (2-3, 3-4, 2-6, and 3-6) is depicted versus the coupling $\beta$, for the decomposition of the transfer entropy.
\label{figS}}}\end{figure}

\begin{figure}[ht!]
\includegraphics[width=10cm]{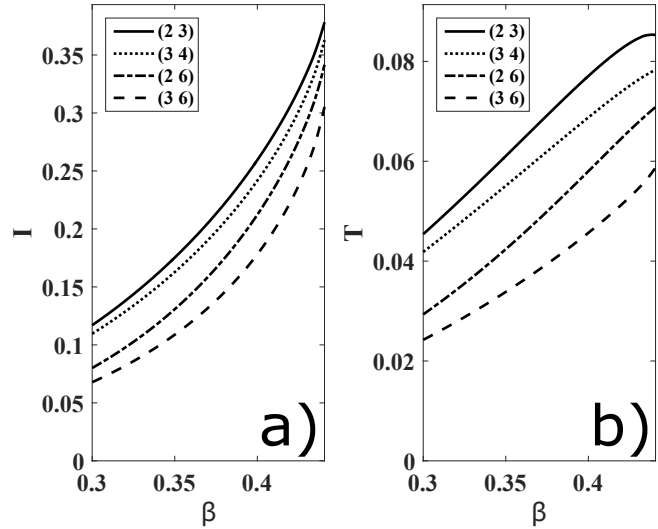}\caption{{\rm (a) The mutual information of four pairs of driving spins (2-3, 3-4, 2-6, and 3-6) is depicted versus the coupling $\beta$. (b) The transfer entropy of four pairs of driving spins (2-3, 3-4, 2-6, and 3-6) is depicted versus the coupling $\beta$.
\label{figg6}}}\end{figure}

It is also worth considering other pairs of driving spins that one can choose from set of nearest and next to nearest neighbors of the target: in figure (\ref{figS}) we depict the synergy, both static and dynamic, for four pairs. We observe that the pair of nearest neighbors $s_3$ and $s_4$ is characterized by a similar behavior to the pair  $s_2$ and $s_3$, but the synergy is lower in this case. This suggests that the knowledge of two opposite spins provides more information, w.r.t. the future of the target, than two nearest neighbor driver spins forming a $\frac{\pi}{2}$ angle with the target: in figure (\ref{figg6}) we depict the mutual information and the transfer entropy for the same sets of drivers, and show that this indeed the case. Intuitively, the reason for this can be ascribed to the fact that when two opposite spins are aligned it is more likely (less surprising) that the target is in the inner of a domain.  It is also worth stressing that figure (\ref{figg6}) shows that $I$ and $T$ do not peak at the same locations as $S^I$ and $S^T$, hence the cause of the peaks of the synergy is not a finite size effect.

\begin{figure}[ht!]
\includegraphics[width=10cm]{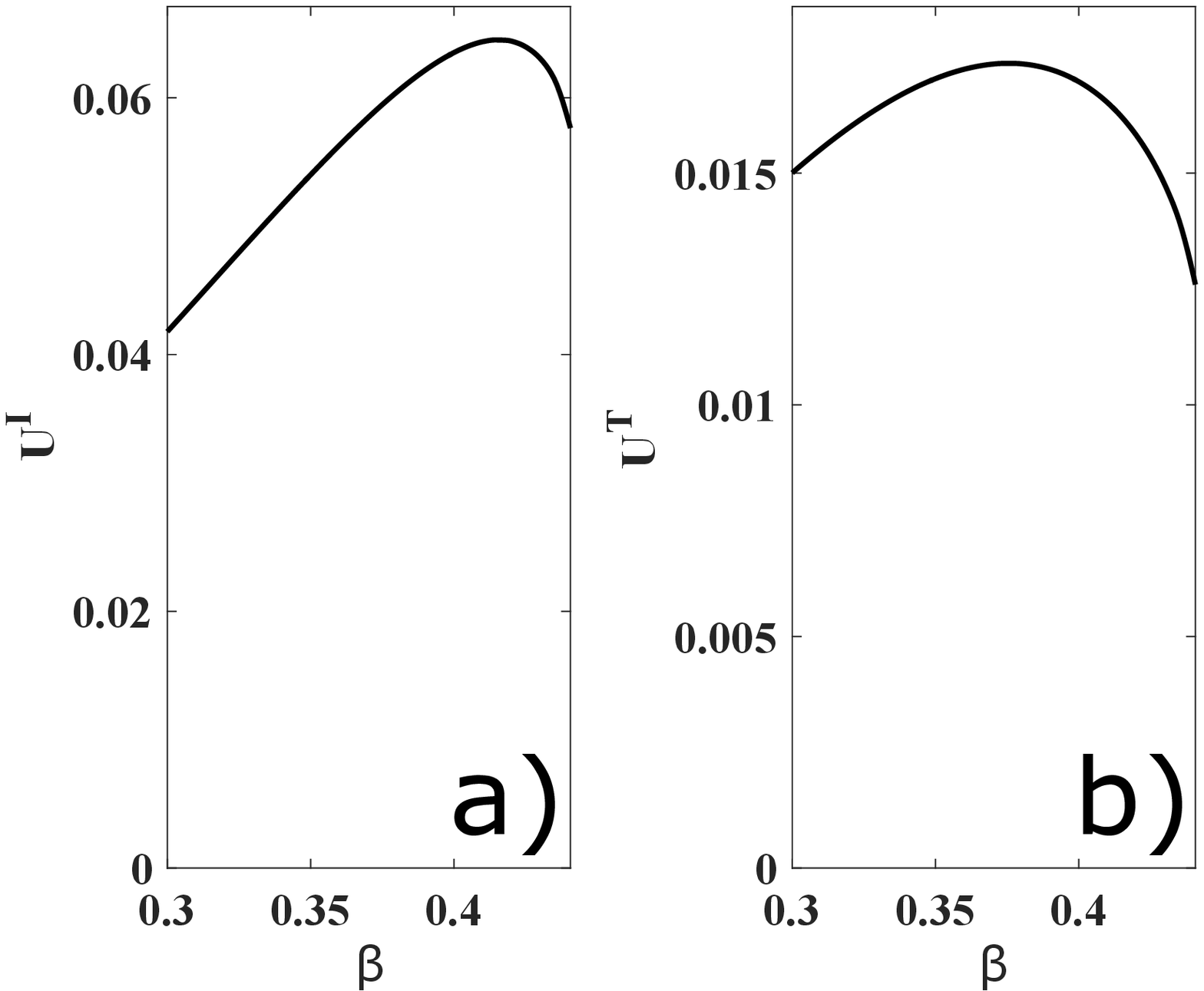}\caption{{\rm The unique information of spin $s_2$ when it is paired with spin $s_6$ in the information decomposition, as a function of the coupling $\beta$. (a) Static case, i.e. mutual information. (b) Dynamic case, i.e. transfer entropy. Note that the same curves hold for the spin $s_3$ when it is paired with $s_6$, as the marginals distributions are the same as in the case of drivers $s_2$ and $s_6$. \label{figu}}
}\end{figure}

Coming back to figure (\ref{figS}), and considering a pair with one nearest neighbor and one next-to-nearest neighbor, 2-6 and 3-6, we note that the interaction of the two drivers diminishes w.r.t. the case of two nearest neighbors. Figure (\ref{figS}) shows that the value of $\beta$ at which the synergy peaks approaches the critical value as the amount of synergy decreases; for the pair 3-6 no peak is observed.  However, in these cases we observe an interesting phenomenon: the unique information from the nearest neighbor driver does not vanish, and peaks in the paramagnetic phase as shown in figure (\ref{figu}). As shown in \cite{matsudo} and \cite{barnettprl2013}, both the pairwise mutual information and the pairwise transfer entropy peak at criticality; however the unique information, obtained subtracting the information redundant with a next-to-the nearest neighbor spin, is non-zero and peaks in the paramagnetic phase. 

Summarizing, we have shown that in the 2D Ising model three variables are sufficient to build reliable precursors of the transition, in the frame of the information
decomposition in a redundant term and a synergistic one, and that this phenomenon is robust w.r.t.  the approach used to perform the information decomposition, indeed we find the same results using two different prescriptions. The redundancy peaks at criticality, while the synergy has a peak in the disordered phase, both in the static case (corresponding to the decomposition of the mutual information)  and in the dynamic case (corresponding to the decomposition of the transfer entropy). The peaks by the dynamic transfer entropy decomposition are far more pronounced than those from the decomposition of mutual information: on one hand this suggests that the study of dynamical influences are more effective to build precursors of transitions, on the other hand the search for the reason of this occurrence deserves further investigation. We have shown that just two variables are required to build the precursor, in comparison with the GTE which requires the knowledge of all the relevant variables. These results also provide an explanation of the fact, noted in \cite{barnettprl2013}, that using just one driving variable one gets quantities peaking at criticality: indeed there are no synergistic components when only one driver is considered. 
Further work could be devoted to extend the present analysis to transitions in networked systems \cite{complex}. 

\begin{acknowledgments}
The computational resources (Stevin Supercomputer Infrastructure) and services used in this work were provided by the VSC (Flemish Supercomputer Center), funded by Ghent University, FWO and the Flemish Government – department EWI. Each realization, with $10^5$ updates, on a 512 2D lattice, took about 760 minutes on a 2GHz CPU. The code for is available at \url{https://github.com/danielemarinazzo/PID_Ising2D}
\end{acknowledgments}


\begin{thebibliography}{99}
\bibitem{barnettprl2013} L. Barnett, J.T. Lizier, and M. Harré, A.K. Seth,and T. Bossomaier,Phys. Rev. Lett. {\bf 111}, 177203 (2013).
\bibitem{schreiber} T. Schreiber, Phys. Rev. Lett. {\bf 85}, 461 (2000).
\bibitem{book_te} T. Bossomaier, L. Barnett, M. Harré, J.T. Lizier, {\it An Introduction to Tranfer Entropy}, Springer 2016.
\bibitem{prokopenko} J. T. Lizier, M. Prokopenko, and A. Y. Zomaya, Chaos {\bf 20}, 037109 (2010).
\bibitem{potts} J. Brown, T. Bossomaier, and L. Barnett  {\it Information Flow in First-Order Potts Model Phase Transition}, preprint arXiv:1810.09607
\bibitem{stock} T. Bossomaier, L. Barnett, A. Steen, M. Harré, S. d'Alessandro,and R. Duncan, Accounting $\&$ Finance {\bf 58} 45-58 (2018)
\bibitem{sootla}S. Sootla, D.O. Theis and R. Vicente, Entropy {\bf 19}, 636 (2017).
\bibitem{MMI} N. Bertschinger, J. Rauh, E. Olbrich, J. Jost, N. Ay, Entropy {\bf 16} 2161 (2014). 
\bibitem{baxter}R. Baxter, {\it Exactly Solved Models in Statistical Mechanics}, Academic Press, London, 1989
\bibitem{matsudo}H. Matsuda, K. Kudo, R. Nakamura, O. Yamakawa, and
T. Murata, Int. J. Theor. Phys. {\bf 35} 839 (1996).
\bibitem{ID} J.T. Lizier, N. Bertschinger, J. Jost, and M. Wibral,  Entropy 2018, 20(4), 307 {\bf 20}, 207 (2018).
\bibitem{harder} M. Harder,C. Salge, and D. Polani, Phys. Rev. E {\bf 87} 012130 (2013). 
\bibitem{griffith} V. Griffith, E.K. Chong, R.G. James, C.J. Ellison, J.P. Crutchfield, Entropy {\bf 16} 1985 (2014).
\bibitem{quax} R. Quax,O. Har-Shemesh, P.M.A. Sloot,  Entropy {\bf 19} 85 (2017).
\bibitem{barrett} A.D. Barrett, Phys. Rev. E {\bf 91}, 052802 (2015).  
\bibitem{complex}
F. Igĺoi, L. Turban, Phys. Rev. E 66, 036140 (2002);
S. Dorogovtsev, A. V. Goltsev, J. F. F. Mendes, Eur. Phys. Journ. B 38, 177 (2004).
\end{thebibliography}
\end{document}